\pdfoutput=1

\documentclass[sigconf]{acmart}

\AtBeginDocument{%
  \providecommand\BibTeX{{%
    \normalfont B\kern-0.5em{\scshape i\kern-0.25em b}\kern-0.8em\TeX}}}

\begin{document}
\fancyhead{}

\copyrightyear{2020}
\acmYear{2020}
\setcopyright{acmcopyright}\acmConference[SIGIR '20]{Proceedings of the 43rd International ACM SIGIR Conference on Research and Development in Information Retrieval}{July 25--30, 2020}{Virtual Event, China}
\acmBooktitle{Proceedings of the 43rd International ACM SIGIR Conference on Research and Development in Information Retrieval (SIGIR '20), July 25--30, 2020, Virtual Event, China}
\acmPrice{15.00}
\acmDOI{10.1145/3397271.3401236}
\acmISBN{978-1-4503-8016-4/20/07}

\title{Training Mixed-Objective Pointing Decoders for Block-Level Optimization in Search Recommendation}

\author{Harsh Kohli}
\affiliation{Microsoft AI \& Research}
\email{hakohl@microsoft.com}


\begin{abstract}
Related or ideal follow-up suggestions to a web query in search engines are often optimized based on several different parameters -- relevance to the original query, diversity, click probability etc.  One or many rankers may be trained to score each suggestion from a candidate pool based on these factors. These scorers are usually pairwise classification tasks where each training example consists of a user query and a single suggestion from the list of candidates. We propose an architecture that takes all candidate suggestions associated with a given query and outputs a suggestion block. We discuss the benefits of such an architecture over traditional approaches and experiment with further enforcing each individual metric through mixed-objective training.
\end{abstract}

\maketitle

\section{Introduction}
Popular search engines often provide related, or potential follow up search recommendations for a user query in the form of features such as Related Searches. Typically, candidate suggestions for these recommendations are mined from past user sessions which are later ranked based on data from user clicks over time. Diversity is maintained either through removal of duplicates or techniques such as maximal marginal relevance (MMR) \cite{Carbonell:1998:UMD:290941.291025} which are employed as a post-processing step. More recent approaches involve training a unified model for direct optimization of diversity \cite{Xia:2015:LMM:2766462.2767710}. In the same vein, we propose a Pointer Network \cite{Vinyals:2015:PN:2969442.2969540} style decoding architecture to model inter-suggestion awareness and optimize on the conditional probabilities of the output sequence of suggestions. This would enable a block-level optimization over all the suggestions as opposed to a pairwise ranking between a query and a suggestion.

Pointer Networks, though originally demonstrated on various geometric combinatorial optimization problems, have been successfully used in NLP tasks such as Machine Comprehension due to their ability to model output sequences with changing vocabularies through a neural attention mechanism capable of pointing back to the input sequence. This application was first proposed by \citet{DBLP:journals/corr/WangJ16a} for the the SQuAD \cite{rajpurkar-etal-2016-squad} question answering task. The boundary model described in the paper consisted of 2 pointers pointing to the start and end words of an answer span in text.

We extend this thought from a word level to a sentence level and describe a novel mechanism to point to query-aware suggestion vectors from a candidate pool. We add multiple reinforcement learning objectives in addition to the standard cross entropy loss to emphasize on the various individual tasks, namely, diversity among suggestions and expected user engagement, and we discuss how the model performs under various settings in the presence or absence of these objectives.

\section{Related Work \& Motivation}

Popular Machine Learning approaches to ranking include RankSVM \cite{article2} and RankBoost \cite{article}. Both of these are pairwise tasks which return confidence scores which are later sorted to decide the relevance between a query and a document. ListNet \cite{listnet}, AdaRank \cite{adarank}, and LambdaMart \cite{lambdamart} were initial approaches to incorporating a list-wise loss function for learning. More recent deep learning approaches such as DSSM \cite{dssm} and CDSSM \cite{cdssm} map each query and document (suggestion, in our case) into a common semantic space. The top k relevant documents can then be determined by cosine similarity of the query and suggestion vectors.

Thus, while traditional machine learning approaches to ranking did include list-wise approaches, modern deep learning architectures again resort to scoring each pair individually and then sorting by these relevance scores. Such architectures lack the inherent ability to diversify among suggestions as similar suggestions would, ideally, be scored similarly and the most relevant suggestions are likely to be of similar intent. For a feature like search recommendation, which aims at solving the users' next-intent through a small number of recommendations, diversifying among the top suggestions is of prime importance. The proposed architecture, to our knowledge, is the first deep learning approach to a list-wise ranking capable of optimizing on both relevance and diversity.

\section{Dataset}

Six months of click logs from a widely used commercial search engine was used as an input stream to our data creation pipeline. The logs were grouped by user query and contained the click rate for each corresponding suggestion from a related searches tab. In order to reduce noise in the dataset, thresholds were placed on the frequency of co-occurrence of a query-suggestion pair as well as minimum number of impressions for a query. Gaussian Mixture Models were used to cluster the suggestions into 8 different buckets. This was done in order to maintain diversity and group together suggestions with similar intents. 

512-dimension embeddings for each suggestion derived from the pretrained Universal Sentence Encoder \cite{DBLP:journals/corr/abs-1803-11175} was used as input to clustering. The encoder uses a Deep Averaging Network \cite{iyyer-etal-2015-deep} over a variety of data sources -- Wikipedia, web news, online discussion forums etc. and is augmented from supervised data from the SNLI dataset \cite{bowman-etal-2015-large} to create pretrained sentence level embeddings that have shown strong performance in several transfer learning tasks. 

After clustering, the suggestion corresponding to the highest click rate was taken from each cluster and the 4 best suggestions (again, determined by clicks) were marked as true labels or the output sequence of suggestions, in our case. The Cluster ID’s and the rank for each suggestion within its cluster was preserved as these would be later used to train the reinforcement learning objectives. About 122k queries were filtered using the above pipeline, of which 100k were used in training and the remaining were divided equally into validation and test sets. The average number of suggestions per query was 63.8.

We take the 4 best suggestions due to UX constraints of the desired feature. The number of clusters $k$ was treated as a hyperparameter to the dataset creation pipeline. Decreasing $k$ leads to greater diversity among suggestions as farther intents may now be grouped together. However, this also means that we run the risk of choosing our final suggestions from less relevant clusters with significantly lower click-rates.  

\section{Model Architecture}

We first define some basic terminology which will be used in later sections. Let $Q = \{Q_1,Q_2,Q_3.....Q_n\}$ be the variable sized set of candidate suggestions for a query $q$. $C = \{C_1,C_2,C_3.....C_n\}$ represents the cluster id's and $R = \{R_1,R_2,R_3.....R_n\}$ represents the rank of each suggestion within its respective cluster when sorted by click rates. $Y = \{Y_1,Y_2,Y_3,Y_4\}$ is the ordered set of 4 output suggestions as described in the previous section and $Y_p = \{Y_{p1},Y_{p2},Y_{p3},Y_{p4}\}$ is the ordered set of suggestions as predicted by our model. Similarly, $C_p = \{C_{p1},C_{p2},C_{p3},C_{p4}\}$ and $R_p = \{R_{p1},R_{p2},R_{p3},R_{p4}\}$ are the clusters and ranks corresponding to the predicted suggestions.

\subsection{Encoder}

Transformer \cite{transformer} based architectures such as BERT \cite{Devlin2019BERTPO} have achieved state-of-the-art results on several NLP tasks such as those on the GLUE \cite{glue} benchmark through transfer learning. Our encoder does a pairwise encoding of a single query and suggestion using the ALBERT \cite{lan2019albert} embedding model which achieves performance close to BERT but scales better due to fewer training parameters and repeating layers. The lower memory footprint helps with keeping the overall architecture within the GPU memory since each training example consists of a query and all its available suggestions. An additional transformation is done after the ALBERT embedding to reduce the output to a lower dimensional vector space. 

\begin{center}
$a_i = \textrm{ALBERT}([Q_i;q])$

$e_i = \textrm{tanh}(W*a_i + b)$
\end{center}

Where $e_i \in{ } \mathbb{R}^{200}$ is the query-aware suggestion embedding for each candidate $Q_i$ in Q. $W \in{ } \mathbb{R}^{768x200}$ and $b$ are trainable parameters to our model. The layers in the ALBERT architecture are kept trainable as well, similar to when fine-tuning on any downstream tasks on GLUE.

\subsection{Decoder}

The dynamic pointing decoder architecture proposed by \citet{DBLP:journals/corr/XiongZS16} used an iterative approach to perform multiple updates on the start and end encodings. This prevented the decoder from prematurely selecting a different but intuitive answer span due to local optima. Candidate questions for a user query may often look semantically very similar, yet have very different click rates based on various subtleties. To make our system robust to these questions, we choose to go with the dynamic pointing decoder architecture, albeit with one small difference. Instead of two pointers pointing to the start and end encoding, we have four pointers that point to the suggestions that form the output sequence of recommendations. 

\begin{center}
    $h_i = \textrm{LSTM}(h_{i-1},[e^{p1}_{i-1};e^{p2}_{i-1};e^{p3}_{i-1};e^{p4}_{i-1}])$
\end{center}

The LSTM Network \cite{Hochreiter:1997:LSM:1246443.1246450} behaves as a state machine which updates its state based on the encodings of the previously selected suggestions - $e^{p1}_{i-1};e^{p2}_{i-1};e^{p3}_{i-1};e^{p4}_{i-1}$. True to the original implementation, we recompute probabilities for the new predicted suggestions using a Highway Network \cite{10.5555/2969442.2969505} with Maxout \cite{Goodfellow:2013:MN:3042817.3043084} activation. Highway Networks facilitate training a network of greater depth, using a gated mechanism to better route information through the various layers. While the highway block used was only a 4-layer network, it still achieved better performance than a similarly sized MLP. Four separate highway networks are trained for each individual pointer.

\begin{center}
    $\alpha_t^k = \textrm{HMN}_{k}(e_{t};h_i;e^{p1}_{i-1};e^{p2}_{i-1};e^{p3}_{i-1};e^{p4}_{i-1})$
\end{center}

Where $t\in\{1,2,3...n\}$ iterates over all the suggestion embeddings from our candidate set and $k\in\{1,2,3,4\}$ represent the 4 HMN networks optimizing towards each individual pointer. Thus, each suggestion receives four different scores corresponding to each individual position and embeddings from previously selected suggestions are attended to when computing the new probabilities. Pointers to the new suggestions are then computed as follows:

\begin{center}
    $p_t^k = \textrm{argmax}(\alpha_1^k, \alpha_2^k, \alpha_3^k....\alpha_n^k)$
\end{center}

The decoding iterations are continued until the output pointers do not change over two iterations or a maximum number of iterations (set to 8) is reached. The sum of cross entropy loss for each iteration of the decoder is optimized, thus incentivizing the network to converge in fewer iterations.

The LSTM cell updates its state based on encodings from all four predicted suggestions and the Highway Maxout architecture takes these as inputs, along with the updated state and each suggestion encoding. Both of these are key elements that enable the architecture to optimize at a block-level and model the interactions between the predicted suggestions and the rest of the candidates.



\section{Mixed-Objective Optimization}

The dynamic pointing decoder was augmented by \citet{DBLP:journals/corr/abs-1711-00106} by introducing an RL-based objective to bridge the disparity between the evaluation metrics and learning objective as defined by the cross entropy loss function. We reason that this is valid for our scenario as well, and describe the various reward functions used in our experiments.

\subsection{Recall or F1 Overlap Reward}

The cross entropy loss penalizes the model on the probabilities of true labels in a ranked order of preference. However, one might contend that ranking between the suggested suggestions might be less important than finding the best four - even if doing so in a jumbled order. This is specially true in our case since we restrict ourselves to a small block of only four suggestions that are immediately visible to the user. Thus, we use recall or overlap between the predicted and output sequence of suggestions. 
\begin{center}
    $R_{overlap} = \frac{|Y \cap Y_p|}{|Y|}$
\end{center}

\subsection{Diversity Reward}

While diversity is intrinsically captured by the output sequence of suggestions, we experiment with further enforcing it through a separate learning objective. This way, the agent gets some reward for picking a diverse set of suggestions even if they do not correspond to the output sequence. We define the diversity loss as the ratio of the number of unique clusters corresponding to the predicted suggestions and the total number of suggestions.
\begin{center}
    $R_{diversity} = \frac{|C_p|}{|Y_p|}$
\end{center}

\subsection{MRR Reward}

We use a third reward function to facilitate a smoother optimization based on click rates. The mean reciprocal ranks (MRR) of the predicted suggestions are used to ensure that good, close suggestions with high (albeit not the highest) click rates are not penalized the same way as decidedly worse suggestion with low user engagement. The objective here is to maximize the cumulative sum of the MRR values of the predicted suggestions.
\begin{center}
    $R_{mrr} = \sum_{i=1}^{4} \frac{1}{R_{pi}}$
\end{center}

\subsection{Loss Computation}

Predicted suggestions are sampled from each time step of the decoder loop and the corresponding rewards $R_{overlap}$, $R_{diversity}$, and $R_{mrr}$ are computed. Based on results by \citet{Greensmith:2004:VRT:1005332.1044710}, we subtract each reward from a strong baseline to help convergence. For this, we use a self-critic \cite{NIPS1999_1786}. In other words, the rewards at each decoding time step are subtracted from the reward of the final predicted suggestions. The RL-loss is the negative of the expected reward at each time step.
\begin{center}
    $l_{rl} = -E_{t}[R_{t} - R_p]$
\end{center}

The gradients for the non-differentiable reward functions were computed based on results by \citet{Sutton:1999:PGM:3009657.3009806} and \citet{Schulman:2015:GEU:2969442.2969633}. The various individual losses are combined using homoscedastic uncertainty as task-dependent weighting \cite{DBLP:journals/corr/KendallGC17}.
\begin{center}
    $loss = \sum_{i=1}^{4} (\frac{1}{2w_{i}^2}l_{i} + \textrm{log}w^{2}_i)$
\end{center}
Where $w_{1}, w_{2}, w_{3}$, and $w_{4}$ are trainable parameters to our model, and  $l_{1}, l_{2}, l_{3}$, and $l_{4}$ correspond to the cross-entropy, overlap, diversity, and MRR losses respectively.

\section{ALBERT + MMR Baseline}

We use MMR, a popular reordering technique for diversity as our baseline. We train a binary classification algorithm, and for the sake of consistency, use the same ALBERT encoder described earlier followed by an output layer. Inputs to the algorithm are now converted to the query-suggestion pairwise format and the outputs are converted to binary indicating whether the suggestion is part of our output sequence. Scores from this classifier are input to MMR along with the cosine similarity between suggestions to find a reordered list. Top 4 suggestions are then picked from this list. 

\section{Experiments \& Results}

Multiple iterations of training were performed with different combinations of rewards in addition to the cross entropy loss as well as the ALBERT baseline with MMR reordering. In addition to the diversity and recall as described above, we log two additional metrics. P@1 is the ratio of the number of times the algorithm correctly predicts the first or highest ranked suggestion. Exact Match (EM) is the ratio of the number of times the algorithm correctly predicts all 4 suggestions in the right order. P@1 is logged to ensure that the network is able to distinguish the most relevant suggestions at par with pairwise approaches and place them at the top. 

\begin{table}[h!]
\begin{center}
 \begin{tabular}{||c c c c c||} 
 \hline
 Objectives & Div Score & Recall & P@1 & EM\\ [0.5ex]  

 \hline\hline
 ce & 0.64193 & 0.44962 & 0.61524 & 0.19416\\ 
 \hline
 ce+f1 & 0.68685 & 0.46585 & 0.63244 & 0.21742\\
 \hline
 ce+f1+mrr & 0.69332 & 0.47637 & 0.62559 & 0.23145\\
 \hline
 ce+f1+div & 0.73774 & 0.49829 & 0.65492 & 0.24555\\
 \hline
 ce+f1+mrr+div & 0.74655 & 0.50168 & 0.66891 & 0.26178\\
 \hline
 mmr($\gamma = 0.6$) & 0.65108 & 0.27654 & 0.62133 & 0.15416\\ 
 \hline
 mmr($\gamma = 1.0$) & 0.30094 & 0.22468 & 0.62133 & 0.11621\\ 
 \hline
 \end{tabular}
 \caption{Test Scores}
\label{table:1}
\end{center}
\end{table}
\vspace*{-\baselineskip}

 Final test metrics are included in Table 1. We see that diversity in the validation set improves significantly with the introduction of the diversity reward in the reinforcement learning objective. The model with all three additional learning objectives performs the best among the various pointer networks on all our evaluation metrics, confirming that a direct optimization of the evaluation metrics helps improve model performance. Interestingly, the P@1 for the pointing decoder architecture is better than the pairwise approach. We reckon this is due to greater number of trainable parameters of the sequential pointing decoder as opposed to a single output layer.

The MMR baseline performs significantly worse on the Recall and Exact Match metrics. Decreasing the $\gamma$ value certainly does help increase diversity. However, the low overlap ratio indicates that while we're choosing diverse suggestions, they are not the optimal or best suggestions available.

\begin{table}[h!]
\begin{center}
 \begin{tabular}{|c|} 
 \hline
thyroid medicine\\ 
 \hline\hline
what do thyroid levels mean?\\
 \hline
what are the symptoms of hyperthyroidism?\\
 \hline
what should i avoid while taking levothyroxine?\\
 \hline
how to treat thyroid problems in men?\\ [1ex] 
 \hline
 \end{tabular}
\end{center}
\end{table}

To better illustrate what the model is doing, we take an example from our test set and the suggestions predicted by our best model. The model selects these questions for the query - thyroid medicine from among 61 candidates. The predicted questions cover an array of diverse, yet often relevant, likely follow-up suggestions which may be of interest to the user - ranging from thyroid levels, symptoms, food to avoid, and treatment. The candidate pool consisted of many suggestions that were semantically similar to each of the questions above - including several close reformulations. Once a question is picked, the architecture learns to ignore the close alternatives in favor of other suggestions, thus catering to a wider span of user intent. Such a result is hard to achieve when training atomic query-suggestion pairs which are isolated from all other suggestions available to them during training. The expected result, in such a scenario, is that most similar suggestions would get similar scores leading to lesser diversity at the top of the ranked list.

\section{Conclusion}

In this paper, we introduce an approach to block-level optimization for search recommendation. Through a re-imagining of a typical recommendation pipeline, we demonstrate how an end-to-end deep learning architecture is able to suggest recommendations in a holistic manner whilst being simultaneously aware of all the other suggestions available to it at any point in time. The sequential nature of optimization preserves a natural ordering among suggestions similar to a list-wise ranking system leveraging state-of-the-art encoding architectures. While the system was trained with the simultaneous optimization of click rates and diversity in mind, it could scale to any multi-suggestion recommendation system particularly when candidate suggestions are not truly independent of one another, or a natural ordering among suggestions is desired.

\bibliographystyle{ACM-Reference-Format}
\bibliography{main}


\end{document}